\documentclass[transmag]{IEEEtran}
\usepackage{latexsym}
\usepackage{graphicx}
\usepackage{amsfonts,amssymb,amsmath}
\usepackage{hyperref}

\usepackage{algorithmic}
\usepackage{textcomp}
\usepackage{xcolor}
\usepackage{subfigure}

\def\BibTeX{{\rm B\kern-.05em{\sc i\kern-.025em b}\kern-.08em T\kern-.1667em\lower.7ex\hbox{E}\kern-.125emX}}
\begin{document}

\title{Personalized Federated Learning for Intelligent IoT Applications: A Cloud-Edge based Framework}

\author{
	Qiong Wu, Kaiwen He and Xu Chen,
	\thanks{The authors are with School of Data and Computer Science, Sun Yat-sen University, Guangzhou 510006, China. }}

\IEEEtitleabstractindextext{\begin{abstract}Internet of Things (IoT) have widely penetrated in different aspects of modern life and many intelligent IoT services and applications are emerging. Recently, federated learning is proposed to train a globally shared model by exploiting a massive amount of user-generated data samples on IoT devices while preventing data leakage. However, the device, statistical and model heterogeneities inherent in the complex IoT environments pose great challenges to traditional federated learning, making it unsuitable to be directly deployed. In this paper we advocate a personalized federated learning framework in a cloud-edge architecture for intelligent IoT applications. To cope with the heterogeneity issues in IoT environments, we investigate emerging personalized federated learning methods which are able to mitigate the negative effects caused by heterogeneities in different aspects. With the power of edge computing, the requirements for fast-processing capacity and low latency in intelligent IoT applications can also be achieved. We finally provide a case study of IoT based human activity recognition to demonstrate the effectiveness of personalized federated learning for intelligent IoT applications.
\end{abstract}

}

\maketitle

\section{INTRODUCTION}
The proliferation of smart devices, mobile networks and computing technology have sparked a new era of Internet of Things (IoT), which is poised to make substantial advances in all aspects of our modern life, including smart healthcare system, intelligent transportation infrastructure, etc \cite{atzori2010internet}. With huge amount of smart devices connected together in IoT, we are able to get access to massive user data to yield insights, train task-specified machine learning models and utimately provide high-quality smart services and products. To reap the benefits of IoT data, the predominant approach is to collect scattered user data to a central cloud for modeling and then transfer the trained model to user devices for task inferences. This kind of approach can be ineffective as data transmission and model transfer will result in high communication cost and latency \cite{zhou2019edge}. Moreover, as the user-sensitive data are required to upload to the remote cloud, it may impose great privacy leakage risk. Under the increasing stringent data privacy protection legislation such as General Data Protection Regulation (GDPR) \cite{voigt2017eu}, the data movement would face unprecedented difficulties. An alternative is to train and update the model at each IoT device with its local data, in isolation from other devices. However, one key impediment of this approach lies in the high resource demand for deploying and training models on IoT devices with limited computational, energy and memory resources. Besides, insufficient data samples and local data shifts will lead to an even worse model.

A sophisticated solution to deal with distributed data training is federated learning which enables to collaboratively train a high-quality shared model by aggregating and averaging locally-computed updates uploaded by IoT devices \cite{yang2019federated}. The primary advantage of this approach is the decoupling of model training from the need for direct access to the training data, and thus federated learning is able to learn a satisfactory global model without compromising user data privacy. Nevertheless, there are three major challenges in the key aspects of federated learning process in the complex IoT environments, making it unsuitable to directly deploy federated learning  in IoT applications.

These three challenges faced by federated learning can be summarized as (1) device heterogeneity, such as varying storage, computational and communication capacities; (2) statistical heterogeneity like the non-IID (a.k.a. non independent and identically distributed) nature of data generated from different devices; (3) model heterogeneity, the situation where different devices want to customize their models adaptive to their application environments. Specifically, resource-constrained IoT devices will be only allowed to train lightweight models under certain network conditions and may further result in high communication cost, stragglers and fault tolerance issues which can not be well handled by traditional federated learning. As federated learning focuses on achieving a high-quality global model by extracting common knowledge of all participating devices, it fails to capture the personal information for each device, resulting in a degraded performance for inference or classification. Furthermore, traditional federated learning requires all participating devices to agree on a common model for collaborative training, which is impractical in realistic complex IoT applications.

To tackle these heterogeneity challenges, one effective way is to perform personalization in device, data and model levels to mitigate heterogeneities and attain high-quality personalized model for each device. Due to its broad application scenarios (e.g., IoT based personalized smart healthcare, smart home services and applications, fine-grained location-aware recommendation services, and on-premise intelligent video analytics), personalized learning has recently attracted great attention \cite{vanhaesebrouck2017decentralized, bellet2017personalized}. We investigate the emerging personalized federated learning approaches which can be the viable alternative to traditional federated learning and summarize them into four categories: federated transfer learning, federated meta learning, federated multi-task learning and federated distillation. These approaches are able to alleviate different kinds of heterogeneity issues in the complex IoT environments and can be promising enabling techniques for many emerging intelligent IoT applications.

In this paper, we propose a synergistic cloud-edge framework named PerFit for personalized federated learning which mitigates the device heterogeneity, statistical heterogeneity and model heterogeneity inherent in IoT applications in a holistic manner. To tackle the high communication and computation cost issues in device heterogeneity, we resort to edge computing which brings the necessary on-demand computing power in the proximity of IoT devices \cite{zhou2019edge}. Therefore, each IoT device can choose to offload its computationally-intensive learning task to the edge which fulfills the requirement for fast-processing capacity and low latency. Besides, edge computing can mitigate privacy concerns by storing the data locally in proximity (e.g., in the smart edge gateway at home for smart home applications) without uploading the data to the remote cloud \cite{lin2017survey}. Furthermore, privacy and security protection techniques such as differential privacy and homomorphic encryption can be adopted to enhance the privacy protection level. For statistical and model heterogeneities, this framework also enables that end devices and edge servers jointly train a global model under the coordination of a central cloud server in a cloud-edge paradigm. After the global model is trained by federated learning, at the device side, different kinds of personalized federated learning approaches can be then adopted to enable personalized model deployments for different devices tailored to their application demands. We further illustrate a representative case study based on a specific application scenario---IoT based activity recognition, which demonstrates the superior performance of PerFit for high accuracy and low communication overhead.

The remainder of this paper is organized as follows. The following section discusses the main challenges of federated learning in IoT environments. To cope with these challenges, we advocate a personalized federated learning framework based on cloud-edge architecture and investigate some emerging solutions to personalization. Then we evaluate the performance of personalized federated learning methods with a motivating study case of human activity recognition. Finally, we conclude the paper.

\section{Main Challenges of Federated Learning in IoT Environments}
\label{SectionChallenges}
In this section, we first elaborate the main challenges and the potential negative effects when using traditional federated learning in IoT environments.

\subsection{Device Heterogeneity}
\label{DH}
There are typically a large number of IoT devices that differ in hardware (CPU, memory), network conditions (3G, 4G, WiFi) and power (battery level) in IoT applications, resulting in diverse computing, storage and communication capacities. Thus, device heterogeneity challenges arise in federated learning, such as high communication cost, stragglers and fault tolerance \cite{smith2017federated}. In federated setting, communication costs are the principal constraints considering the fact that IoT devices are frequently offline or on slow or expensive connections \cite{mcmahan2016communication}. In the federated learning process performing a synchronous update, the devices with limited computing capacity could become stragglers as they take much longer to report their model updates than other devices in the same round. Moreover, participating devices may drop out the learning process due to poor connectivity and energy constraints, causing a negative effect on federated learning. As the stragglers and faults issues are very prevalent due to the device heterogeneity in complex IoT environments, it is of great significance to address the practical issues of heterogeneous device communication and computation resources in federated learning setting.

\subsection{Statistical Heterogeneity}
Consider a supervised task with features $x$ and labels $y$, the local data distribution of user $i$ can be represented as $\mathcal{P}_{i}(x,y)$. Due to users' different usage environments and patterns, the personally-generated data $(x,y)$ from different devices may naturally exhibit the kind of non-IID distributions. As $\mathcal{P}_{i}(x,y) = \mathcal{P}_{i}(y|x)\mathcal{P}_{i}(x) = \mathcal{P}_{i}(x|y)\mathcal{P}_{i}(y)$, user data can be non-IID in many forms, such as feature distribution skew, label distribution skew and concept shift \cite{kairouz2019advances}. For example, in healthcare applications, the distributions of users' activity data differ greatly according to users' diverse physical characteristics and behavioral habits (feature distribution skew). Moreover, the number of data samples across devices may vary significantly \cite{xu2019federated}. This kind of statistical heterogeneity is pervasive in complex IoT environments. To address this heterogeneity challenge, the canonical federated learning approach, FederatedAveraging (FedAvg), is demonstrated to be able to work with certain non-IID data. However, FedAvg may lead to a severely degraded performance when facing highly skewed data distributions. Specifically, on the one hand, non-IID data will result in weight divergence between federated learning process and the traditional centralized training process, which indicates that Fedvg will finally obtain a worse model than centralized methods and thus result in poor performance \cite{zhao2018federated}. On the other hand,  FedAvg only learns the coarse features from IoT devices, while fails in learning the fine-grained information on a particular device.

\subsection{Model Heterogeneity}
In the original federated learning framework, participating devices have to agree on a particular architecture of the training model so that the global model can be effectively obtained by aggregating the model weights gathered from local models. However, in practical IoT applications, different devices want to craft their own models adaptive to their application environments and resource constraints (i.e., computing capacity). And they may be not willing to share the model details due to privacy concerns. As a consequence, the model architectures from different local models exhibit various shapes, making it impossible to perform naive aggregation by traditional federated learning \cite{li2019federated}. In this case, the problem of model heterogeneity turns to become how to enable a deep network to understand the knowledge of others without sharing data or model details. Model heterogeneity inherent in IoT environments has attracted considerable research attention due to its practical significance for intelligent IoT applications.

\begin{figure*}[tbp]
	\centering
	\includegraphics[width=0.9\linewidth]{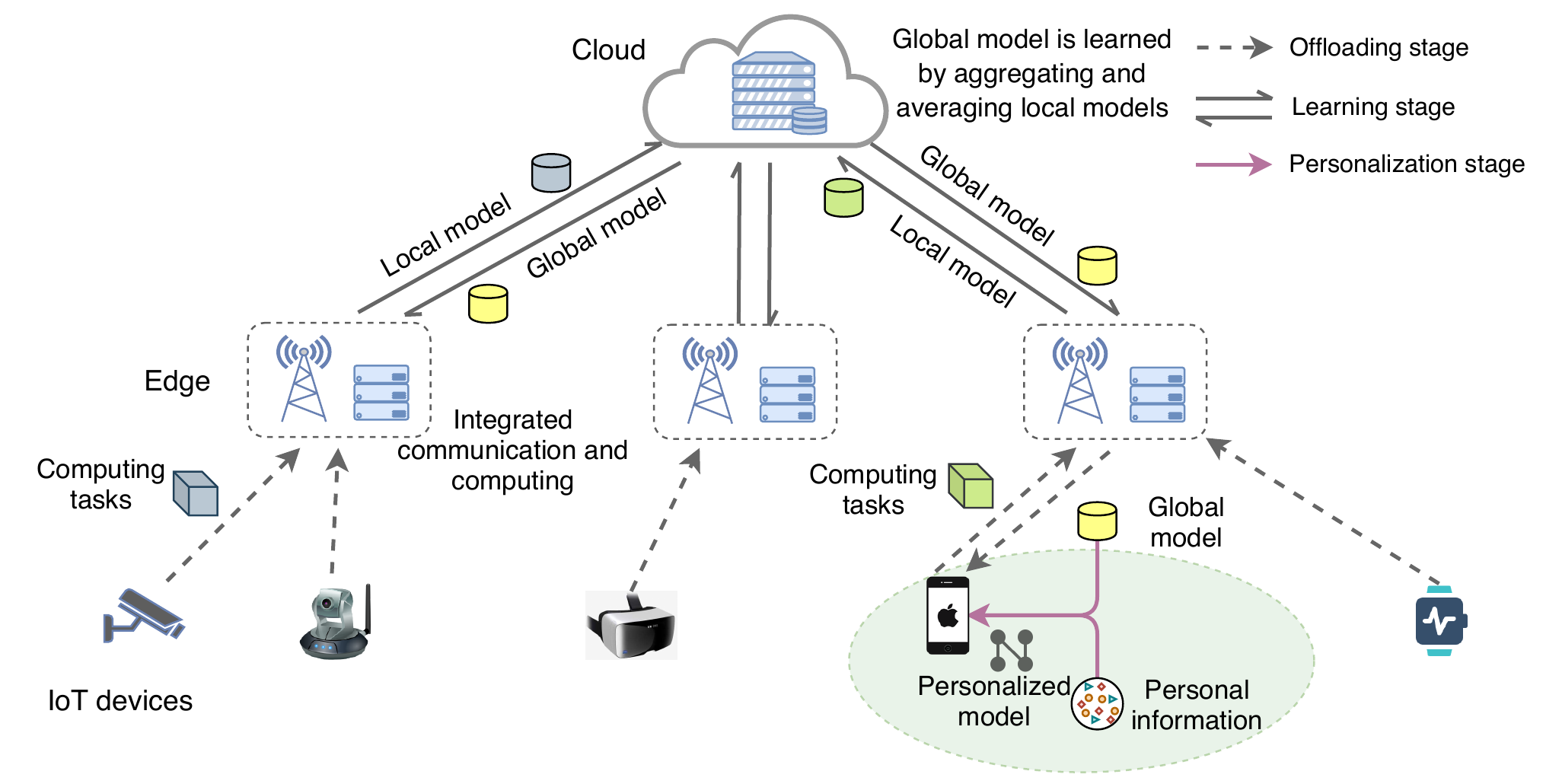}
	\caption{The personalized federated learning framework for intelligent IoT applications, which supports flexible selection of personalized federated learning approaches.}
	\label{fig1}
\end{figure*}

\section{Cloud-Edge Framework for Personalized Federated Learning}
As elaborated in Section \ref{SectionChallenges}, there exist device heterogeneity, statistical heterogeneity and model heterogeneity in IoT applications, which poses great challenges to traditional federated learning. An effective solution for addressing those heterogeneity issues can boil down to personalization. By devising and leveraging more advanced federated learning methods, we aim to enable the great flexibility such that individual devices can craft their own personalized models to meet their resource and application requirements and meanwhile enjoy the benefit from federated learning for collective knowledge sharing.

In this paper, we advocate a personalized federated learning framework for intelligent IoT applications to tackle the heterogeneity challenges in a holistic manner. As depicted in Fig. \ref{fig1}, our proposed PerFit framework adopts a cloud-edge architecture, which brings necessary on-demand edge computing power in the proximity of IoT devices. Therefore, each IoT device can choose to offload its intensive computing tasks to the edge (i.e., edge gateway at home, edge server at office, or 5G MEC server outdoors) via the wireless connections, thus the requirements for high processing efficiency and low latency of IoT applications can be fulfilled.

To support collaborative learning for intelligent IoT applications, federated learning (FL) is then adopted between end devices, edge servers and the remote cloud, which enables to jointly train a shared global model by aggregating locally-computed models from the IoT users at the edge while keeping all the sensitive data on device. To tackle the heterogeneity issues, we will further carry out personalization and adopt some personalized federated learning methods to fine tune the learning model for each individual device.

Specifically, the collaborative learning process in PerFit mainly consists of the following three stages as depicted in Fig. \ref{fig1}:
\begin{itemize}		
	\item \textbf{Offloading stage.} When the edge is trustworthy (e.g., edge gateway at home), the IoT device user can offload its whole learning model and data samples to the edge for fast computation. Otherwise, the device user will carry out model partitioning by keeping the input layers and its data samples locally on its device and offloading the remaining model layers to the edge for device-edge collaborative computing \cite{li2019edge}.
	
	\item \textbf{Learning stage.} The device and the edge collaboratively compute the local model based on personal data samples and then transmit the local model information to the cloud server. The cloud server aggregates local model information submitted by participating edges and averages them into a global model to send back to edges. Such model information exchanging process repeats until it converges after a certain number of iterations. Thus a high-quality global model can be achieved and then transmitted to the edges for further personalization.
	
	\item \textbf{Personalization stage.} To capture the specific personal characteristics and requirements, each device will train a personalized model based on global model information and its own personal information (i.e., local data). The specific learning operations at this stage depend on the adopted personalized federated learning mechanism which will be elaborated in next section.	
\end{itemize}

\begin{figure*}[!t]
	\centering
	\subfigure[The whole trained global model in the cloud server is transferred to the device for personalization with its local data.]{
		\includegraphics[width=0.44\linewidth]{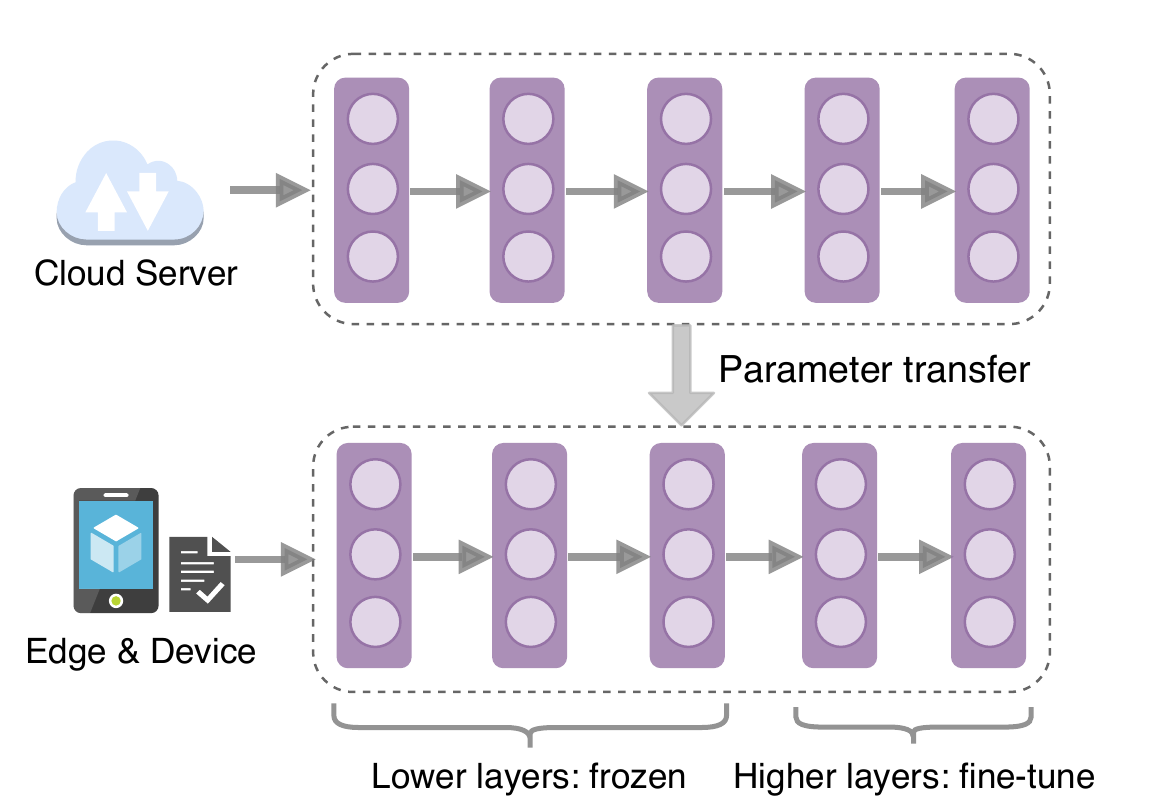}
	}
	\subfigure[The device model is combined with the part of model transferred from cloud server and the personalization layers owned by users locally.]{
		\includegraphics[width=0.45\linewidth]{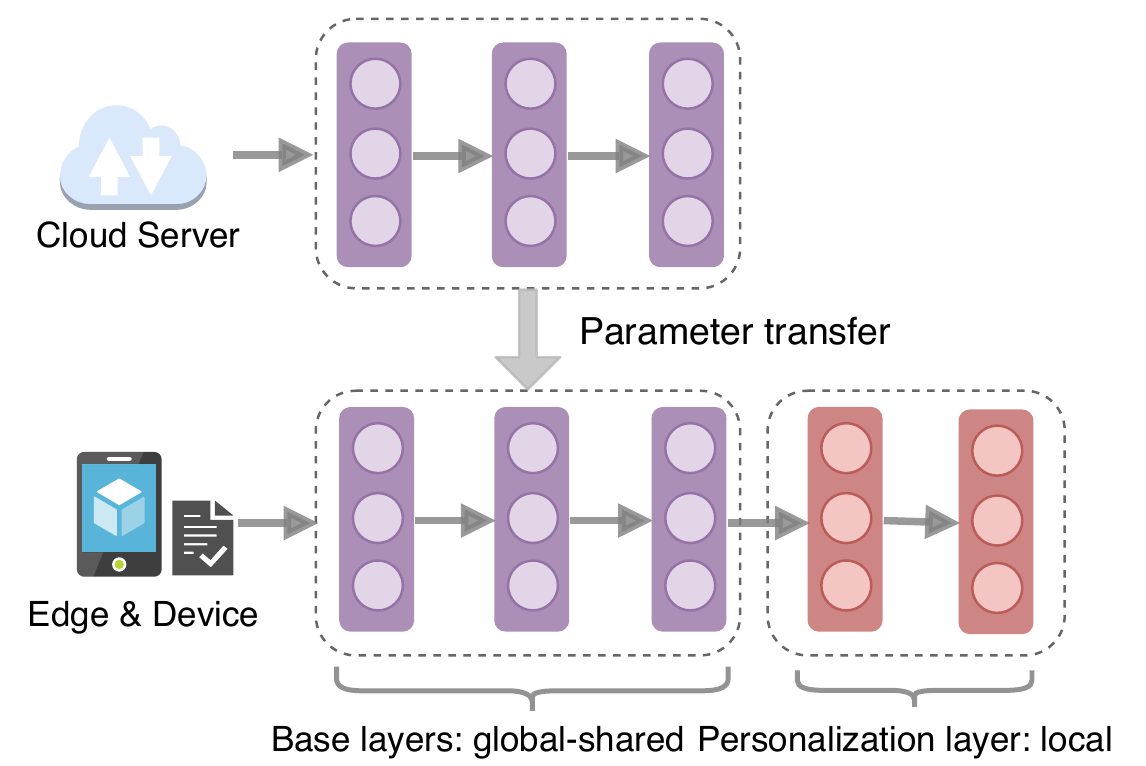}
	} 	
	\caption{Federated transfer learning.}
	\label{fig:user5}
\end{figure*}

The proposed PerFit framework leverages edge computing to augment the computing capability of individual devices via computation offloading to mitigate the straggle effect. If we further conduct local model aggregation at the edge server, it also helps to reduce the communication overhead by avoiding massive devices to directly communicate with the cloud server over the expensive backbone network bandwidth \cite{luo2020hfel}. Moreover, by performing personalization, we can deploy lightweight personalized models at some resource-limited devices (e.g., by model pruning or transfer learning). These would help to mitigate the device heterogeneity in communication and computation resources. Also, the statistical heterogeneity and model heterogeneity can be well supported, since we can leverage personalized models and mechanisms for different individual devices tailored to their local data characteristics, application requirements and deployment environments.

Note that the adopted personalized federated learning mechanism will be the core of the collaborative learning in PerFit, which also determines the exchanging model information between the cloud server and the edges. For example, it is also allowed to transmit only part of the model parameters due to the specific setting of federated transfer learning as we will discuss in the coming section. If facing the situation where different models are trained on different IoT devices, the output class probabilities of local models can be encapsulated as its local information to send to the cloud server via federated distillation approaches. PerFit is flexible to integrate with many kinds of personalized federated methods by exchanging different kinds of model information between the edges and cloud accordingly. By addressing the heterogeneity issues inherent in the complex IoT environments and ensuring user privacy by default, PerFit can be ideal for large-scale practical deployment.

\section{Personalized Federated Learning Mechanisms}
In this section, we review and elaborate several key personalized federated learning mechanisms that can be integrated with PerFit framework for intelligent IoT applications. These personalized federated learning schemes can be categorised by federated transfer learning, federated meta learning, federated multi-task learning and federated distillation, which will be elaborated as follows.

\subsection{Federated Transfer Learning}
Transfer learning \cite{pan2009survey} aims at transferring knowledge (i.e., the trained model parameters) from a source domain to a target domain. In the setting of federated learning, the domains are often different but related, which makes knowledge transfer possible. The basic idea of federated transfer learning is to transfer the globally-shared model to distributed IoT devices for further personalization in order to mitigate the statistical heterogeneity (non-IID data distributions) inherent in federated learning. Considering the architecture of deep neural networks and communication overload, there are two main approaches to perform personalization via federated transfer learning.

Chen et al. \cite{chen2019fedhealth} first train a global model through traditional federated learning and then transfer the global trained model back to each device. Accordingly, each device is able to build personalized model by refining the global model with its local data. To reduce the training overhead, only model parameters of specified layers will be fine-tuned instead of retraining whole model. As presented in Fig. 2 (a), model parameters in lower layers of global model can be transferred and reused directly for local model as lower layers of deep networks focus on learning common and low-level features. While the model parameters in higher layers should be fine-tuned with local data as they learn more specific features tailored to current device. Besides, Feng et al. \cite{feng2020pmf} design two personal adaptors (personal bias, personal filter) for higher layers in user's local model which can be fine-tuned with personal information.

Arivazhagan et al. \cite{arivazhagan2019federated} propose FedPer which takes a different way to perform personalization through federated transfer learning. FedPer advocates viewing deep learning models as \textit{base + personalization} layers as illustrated in Fig. 2 (b). Base layers act as the shared layers which are trained in a collaborative manner using the existing federated learning approach (i.e., FedAvg method). While the personalization layers are trained locally thereby enabling to capture personal information of IoT devices. In this way, after the federated training process, the globally-shared base layers can be transferred to participating IoT devices for constituting their own personalized deep learning models with their unique personalization layers. Thus, FedPer is able to capture the fine-grained information on a particular device for superior personalized inference or classification, and address the statistical heterogeneity to some extent. Besides, by uploading and aggregating only part of the models, FedPer requires less computation and communication overhead, which is essential in IoT environments.

Note that subject to the computing resource constraint of the device, model pruning and compression techniques can be further leveraged to achieve the lightweight model deployment after the personalized model is obtained.

\subsection{Federated Meta Learning}
Federated learning in IoT environments generally faces statistical heterogeneity such as non-IID and unbalanced data distributions, which makes it challenging to ensure a high-quality performance for each participating IoT devices. To tackle this problem, some researchers concentrate on improving FedAvg algorithm by leveraging the personalization power of meta learning. In meta learning, the model is trained by a meta-learner which is able to learn on a large number of similar tasks and the goal of the trained model is to quickly adapt to a new similar task from a small amount of new data \cite{finn2017model}. By regarding the similar tasks in meta learning as the personalized models for the devices, it is a natural choice to integrate federated learning with meta learning to achieve personalization through collaborative learning.

Jiang et al. \cite{jiang2019improving} propose a novel modification of FedAvg algorithm named Personalized FedAvg by introducing a fine-tuning stage using model agnostic meta learning (MAML), a representative gradient-based meta learning algorithm. Thus, the global model trained by federated learning can be personalized to capture the fine-grained information for individual devices, which results in an enhanced performance for each IoT device. MAML is flexible to combine with any model representation that is amenable to gradient-based training. Besides, it can learn and adapt quickly from only a few data samples.

Since the federated meta learning approach often utilizes complicated training algorithms, it has higher implementation complexity than the federated transfer learning approach. Nevertheless, the learned model by federated meta learning is more robust and can be very useful for those devices with very few data samples.

\subsection{Federated Multi-Task Learning}
\begin{figure}[tbp]
	\centering
	\includegraphics[width=0.95\linewidth]{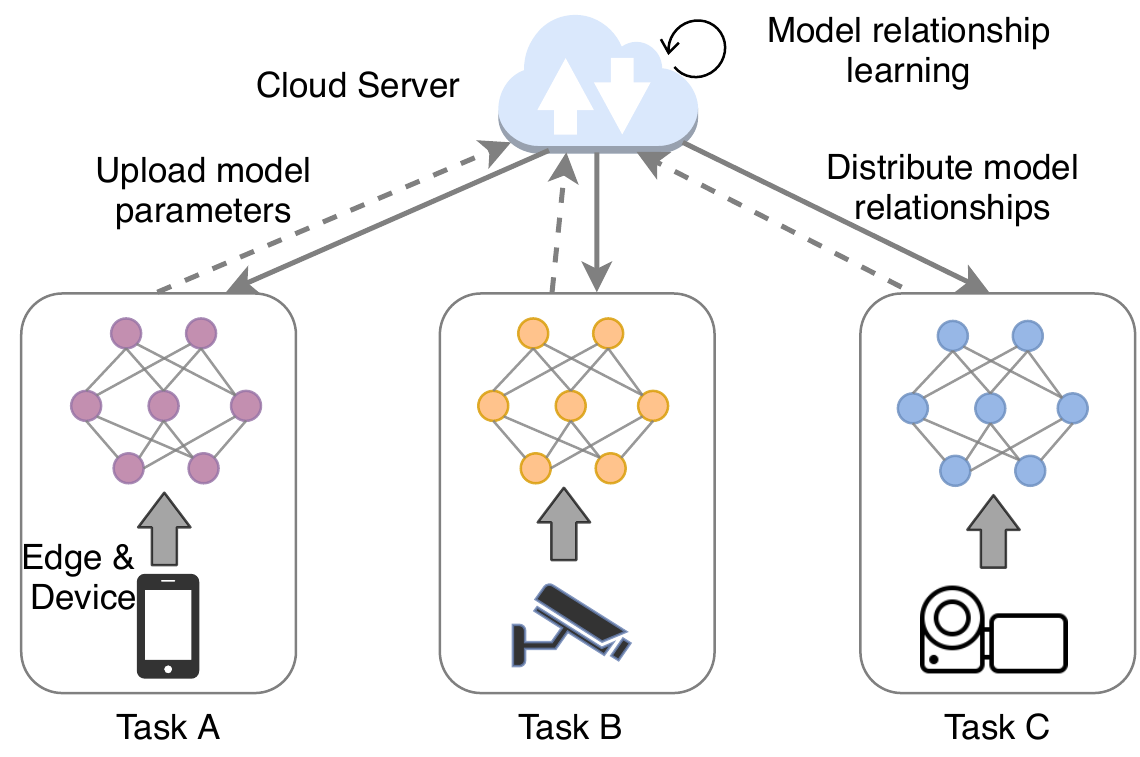}
	\caption{Federated multi-task learning.}
	\label{fig3}
\end{figure}

In general, federated transfer learning and federated meta learning aim to learn a shared model of the same or similar tasks across the IoT devices with fine-tuned personalization. Along a different line, federated multi-task learning aims at learning distinct tasks for different devices simultaneously and tries to capture the model relationships amongst them without privacy risk \cite{corinzia2019variational}. Through model relationships, the model of each device may be able to reap other device's information. Moreover, the model learned for each device is always personalized. As shown in Fig. \ref{fig3}, in the training process of federated multi-task learning, the cloud server learns the model relationships amongst multiple learning tasks based on the uploaded model parameters by IoT devices. And then each device can update its own model parameters with its local data and current model relationships. Through the alternating optimization of model relationships in the cloud server and model parameters for each task, federated multi-task learning enables participating IoT deivices to collaboratively train their local models so as to mitigate statistical heterogeneity and obtain high-quality personalized models.

Smith et al. \cite{smith2017federated} develop a distributed optimization method MOCHA through a federated multi-task learning framework. For high communication cost, MOCHA allows the flexibility of computation which yields direct benefits for communication as performing additional local computation will result in fewer communication rounds in federated settings. To mitigate stragglers, the authors propose to approximately compute the local updates for devices with limited computing resources. Besides, asynchronous updating scheme is also an alternative approach for straggler avoidance. Furthermore, by allowing participating devices periodically dropping out, MOCHA is robust to fault tolerance. As device heterogeneity inherent in complex IoT environments is critical to the performance of federated learning, federated multi-task learning is of great significance for intelligent IoT applications. Nevertheless, as federated multi-task learning produces one model per task, it requires that all clients (e.g., IoT devices) participate in every iteration which is impractical in IoT applications. To tackle this issue, we believe that cluster-based federated multi-task learning is a promising direction in research.

\subsection{Federated Distillation}
In original federated learning framework, all clients (e.g., participating edges and devices) have to agree on a particular architecture of the model trained on both the global server and local clients. However, in some realistic business setting, like healthcare and finance, each participant would have capacity and desire to design its own unique model, and may not be willing to share the model details due to privacy and intellectual property concerns. This kind of model heterogeneity poses new challenge to traditional federated learning.

\begin{figure}[tbp]
	\centering
	\includegraphics[width=0.99\linewidth]{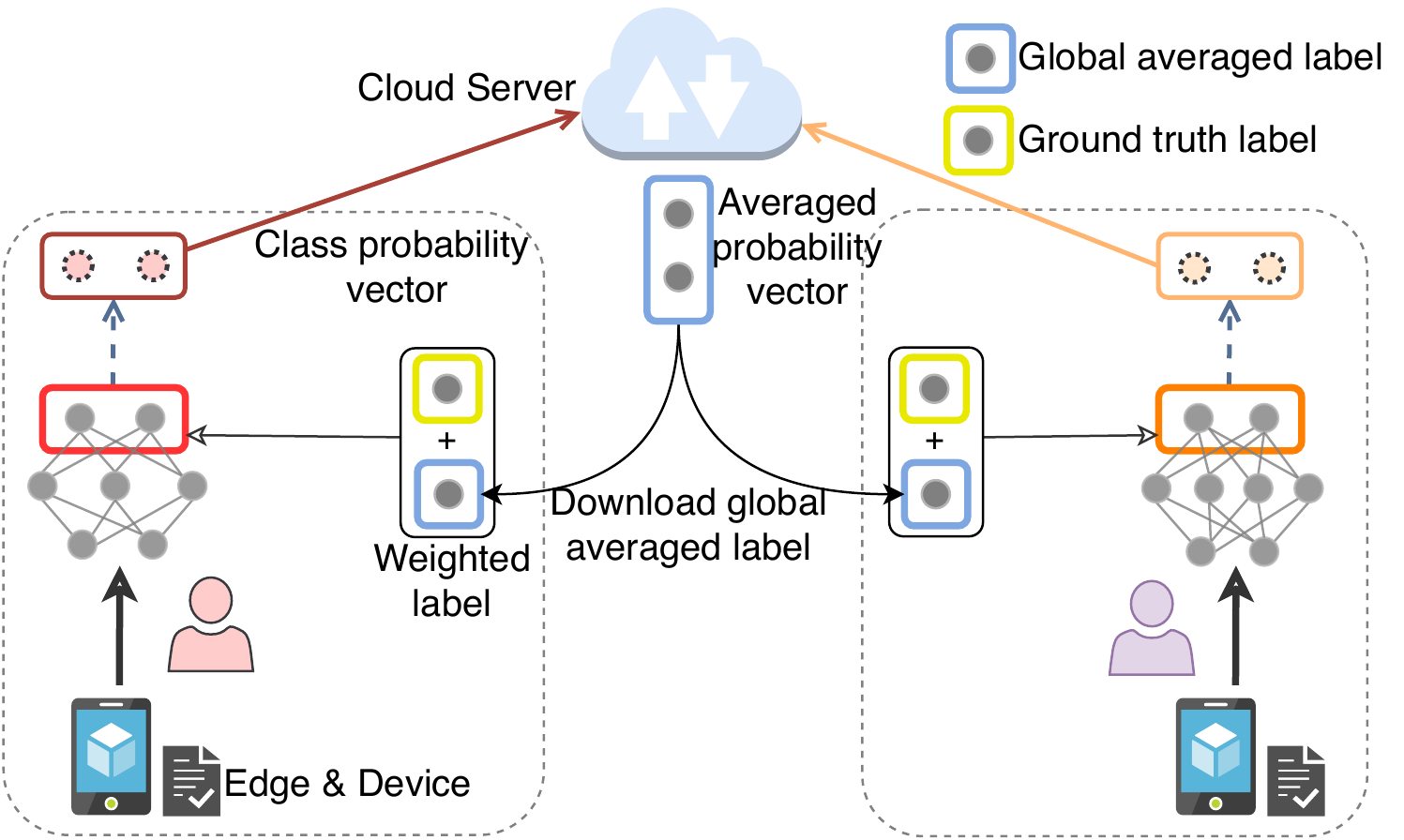}
	\caption{Federated distillation.}
	\label{fig4}
\end{figure}

To tackle this challenge, Li et al. \cite{li2019fedmd} propose FedMD, a new federated learning framework that enables participants to independently design their own models by leveraging the power of knowledge distillation. In FedMD, each client needs to translate its learned knowledge to a standard format which can be understood by others without sharing data and model architecture. And then a central server collects these knowledges to compute a consensus which will be further distributed to the participating clients. The knowledge translation step can be implemented by knowledge distillation, for example, using the class probabilities produced by client model as the standard format as shown in Fig. \ref{fig4}. In this way, the cloud server aggregates and averages the class probabilities for each data sample and then distributes to clients to guide their updates. Jeong et al. \cite{jeong2018communication} propose federated distillation where each client treats itself as a student and sees the mean model output of all the other clients as its teacher's output. The teacher-student output difference provides the learning direction for the student. Here it is worthnoting that, to operate knowledge distillation in federated learning, a public dataset is required because the teacher and student outputs should be evaluated using an identical training data sample. Moreover, federated distillation can significantly reduce the communication cost as it exchanges not the model parameters but the model outputs \cite{ahn2019wireless}.

\subsection{Data Augmentation}
As user's personally-generated data naturally exhibits the kind of highly-skewed and non-IID distribution which may greatly degrade the model performance, there are emerging works focusing on data augmentation to facilitate personalized federated learning. Zhao et al. \cite{zhao2018federated} propose a data-sharing strategy by distributing a small amount of global data containing a uniform distribution over classes from the cloud to the edge clients. In this way, the highly-unbalanced distribution of client data can be alleviated to some extent and then the model performance of personalization can be improved. However, directly distributing the global data to edge clients will impose great privacy leakage risk, this approach is required to make a trade-off between data privacy protection and performance improvement. Moreover, the distribution difference between global shared data and user's local data can also bring performance degradation.

To rectify the unbalanced and non-IID local dataset without compromising user privacy, some over-sampling techniques and deep learning approaches with generative ability are adopted. For example, Jeong et al. \cite{jeong2018communication} propose federated augmentation (FAug), where each client collectively trains a generative model, and thereby augments its local data towards yielding an IID dataset. Specifically, each edge client recognizes the labels being lacking in its data samples, referred to as target labels, and then uploads few seed data samples of these target labels to the server. The server oversamples the uploaded seed data samples and then trains a generative adversarial network (GAN). Finally, each device can download the trained GAN's generator to replenish its target labels until reaching a balanced dataset. With data augmentation, each client can train a more personalized and accurate model for classification or inference based on the generated balanced dataset. It is worthnoting that the server in FAug should be trustworthy so that users are willing to upload their personal data.

\section{Case Study}
In this section, we first describe the experiment settings and then evaluate different personalized federated learning approaches with different kinds of heterogeneities in terms of accuracy and comminication size.

\subsection{Dataset Description and Implementation Details}
In the experiments, we focus on human activity recognition task based on a publicly accessible dataset called MobiAct \cite{vavoulas2016mobiact}. Each volunteer participating in the generation of MobiAct dataset wears a Samsung Galaxy S3 smartphone with accelerometer and gyroscope sensors. The tri-axial linear accelerometer and angular velocity signals are recorded by embedded sensors while volunteers perform predefined activities. We use an 1-second sliding window for feature extraction since one second is enough to perform an activity. There are ten kinds of activities recorded in MobiAct, such as walking, stairs up/down, falls, jumping, jogging, step in a car, etc. To practically mimic the environment of federated learning, we randomly select 30 volunteers and regard them as different clients. For each client, we take a random number of samples for each activity and finally, each client has 480 samples for model training. In this way, the personal data of different clients may exhibit the kind of non-IID distributions (statistical heterogeneity). The test data for each client is composed of 160 samples under a balanced distribution.

In order to meet the needs of different clients for customizing their own models (model heterogeneity) in IoT applications, we design two kinds of models for training on the clients: 1) a Multi-Layer Perceptron network composed of three fully-connected layers with 400, 100 and 10 neural units (521,510 total parameters), which we refer to as the 3NN, 2) a convolutional neural network (CNN) with three $3 \times 3$ convolutional layers (the first with 32 channels, the second with 16, the last with 8, each of the first two layers followed by a $2 \times 2$ max-pooling layer), a fully-connected layer with 128 units and $ReLu$ activation, and a final $Softmax$ output layer (33,698 total parameters). Cross-entropy loss and Stochastic Gradient Descent (SGD) optimizer with a learning rate of 0.01 are used for the training of both 3NN and CNN.

\begin{figure*}[!t]
	\centering
	\subfigure[Test Accuracy]{
		\includegraphics[width=0.486\linewidth]{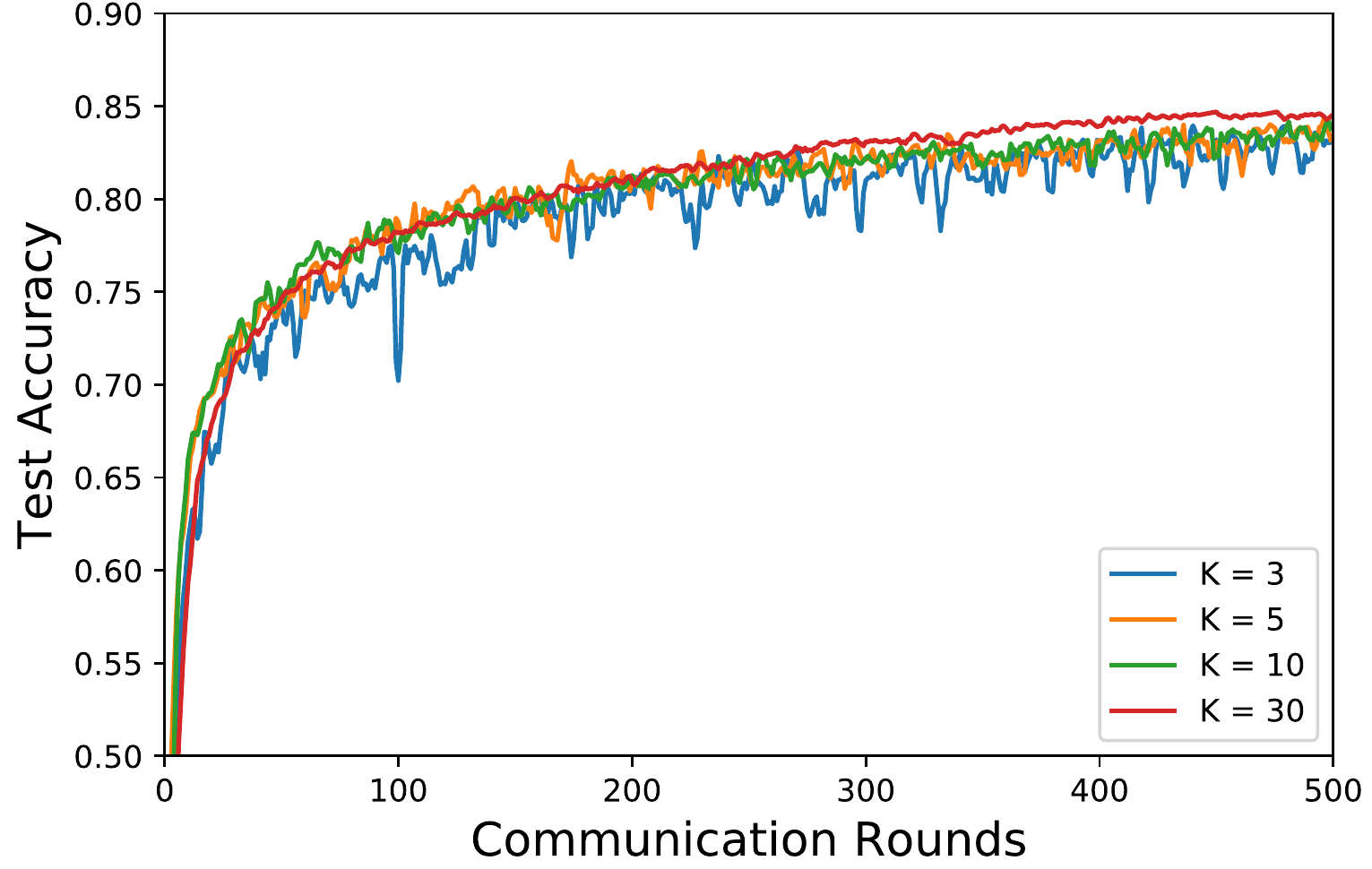}
	}
	\subfigure[Time Cost]{
		\includegraphics[width=0.478\linewidth]{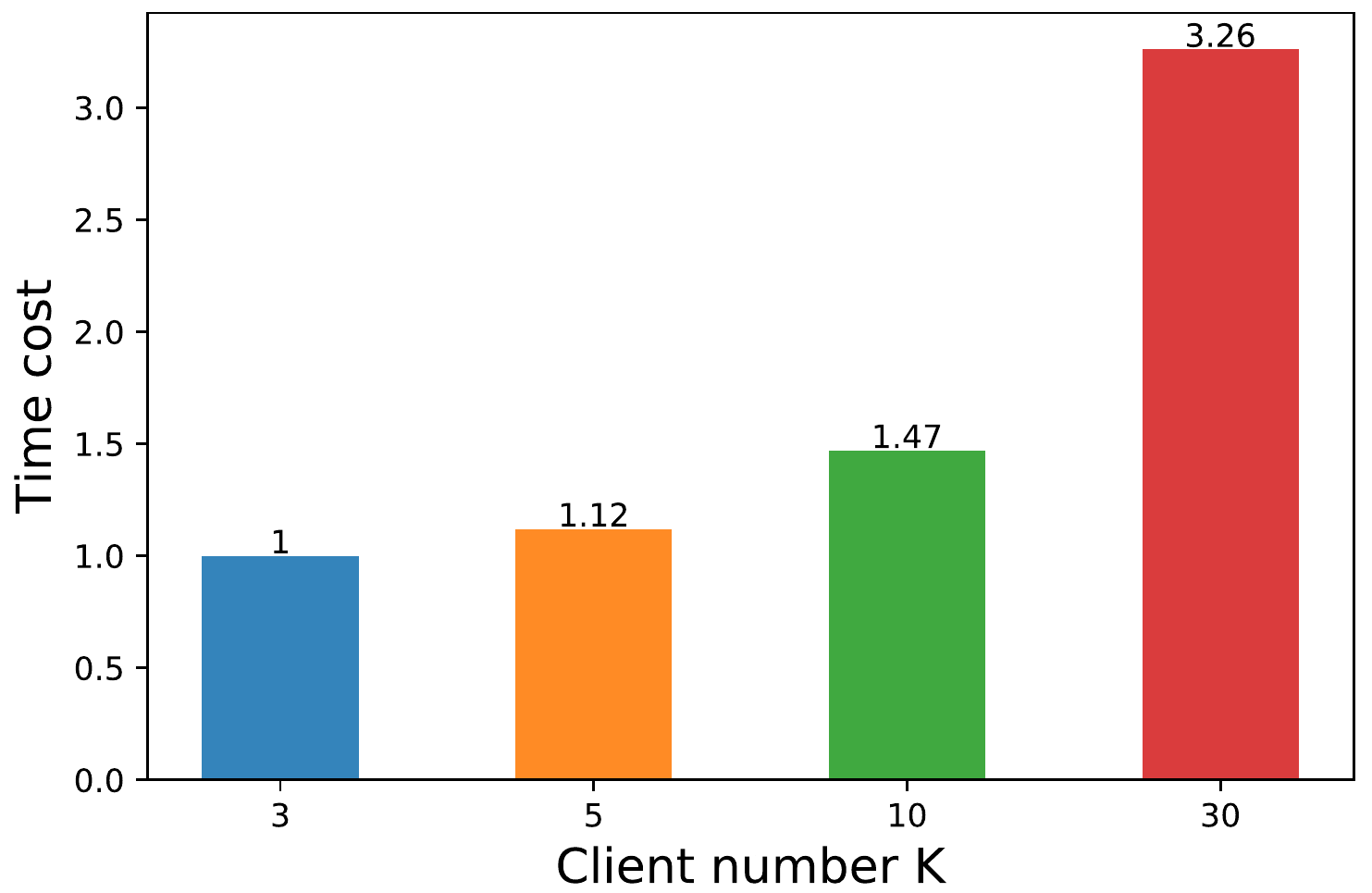}
	} 	
	\caption{The test accuracy and time cost under different number of participating clients in each commuication round. We choose K = 5 by making a trade off between the stability and the efficiency of the learning algorithm.}
	\label{KBE}
\end{figure*}

\subsection{Experimental results}
\subsubsection{Comparing Methods:} We compare the performance of personalized federated learning with both centralized scheme and traditional federated learning. For centralized methods, we adopt the widely-used machine learning approaches in human activity recognition task such as support vector machine (SVM) \cite{ward2016towards}, k-nearest neighbor (kNN) \cite{he2016smart} and random forest (RF) \cite{yuan2014power}. Besides, centralized 3NN (c3NN) and centralized CNN (cCNN) are also used for comparison. As centralized approaches require a large amount of data, we collect all the training data of 30 users for model learning. In traditional federated settings, each client trains a local model (e.g., 3NN or CNN in our experiment) with its personal-generated data. FedAvg method \cite{mcmahan2016communication}, which aggregates local model updates on each client and then sends them to a cloud server that performs model averaging in an iterative way, is applied to train the global model. Then, the well-trained global model in the cloud is directly distributed to clients for human activity recognition. As for personalized FL, we study the performance of the two widely-adopted approaches: federated transfer learning (FTL) and federated distillation (FD). For FTL, each client will fine tune the model downloaded from the cloud server with its personal data. While in FD, each client can customize its own model according to its own requirements. Note that each client is able to offload its learning task from its device to the edge in proximity (e.g., edge gateway at home) for fast computation in our cloud-edge paradigm.

\subsubsection{Performance Evaluation:} As elaborated in Section \ref{DH}, due to the device heterogeneity (communication and computing resources constraints of IoT devices), there are only a few clients participating in the global model learning in each communication round. Thus, we first experiment with the number of participating clients $K$ in each round. We set $K$ equal to 3, 5, 10 and 30, which means that $\frac{1}{10}$, $\frac{1}{6}$, $\frac{1}{3}$ and $100\%$ of users participating in the federated learning process in each communication round. As depicted in Fig. \ref{KBE}(a), for all values of $K$, the test accuracy improves with the number of communication rounds increases and the test accuracies are similar when the training process converges. However, when $K$ is small, the learning curve exists erratic fluctuation to some extent. As $K$ increases, the learning curve becomes smoother and smoother. Although the test accuracies are similar, the training time for each value of $K$ varies dramatically as demonstrated in Fig. \ref{KBE} (b).  For example, the training time for $K = 30$ is $3.26$ times longer than that in the $K = 3$ case. We make a trade-off between the stability and the efficiency for the training process and fix $K = 5$ for the following experiments. For each method, we compute the average of test accuracy by repeating the training and prediction processes five times.

\begin{figure*}[tbp]
	\centering
	\includegraphics[width=0.9\linewidth]{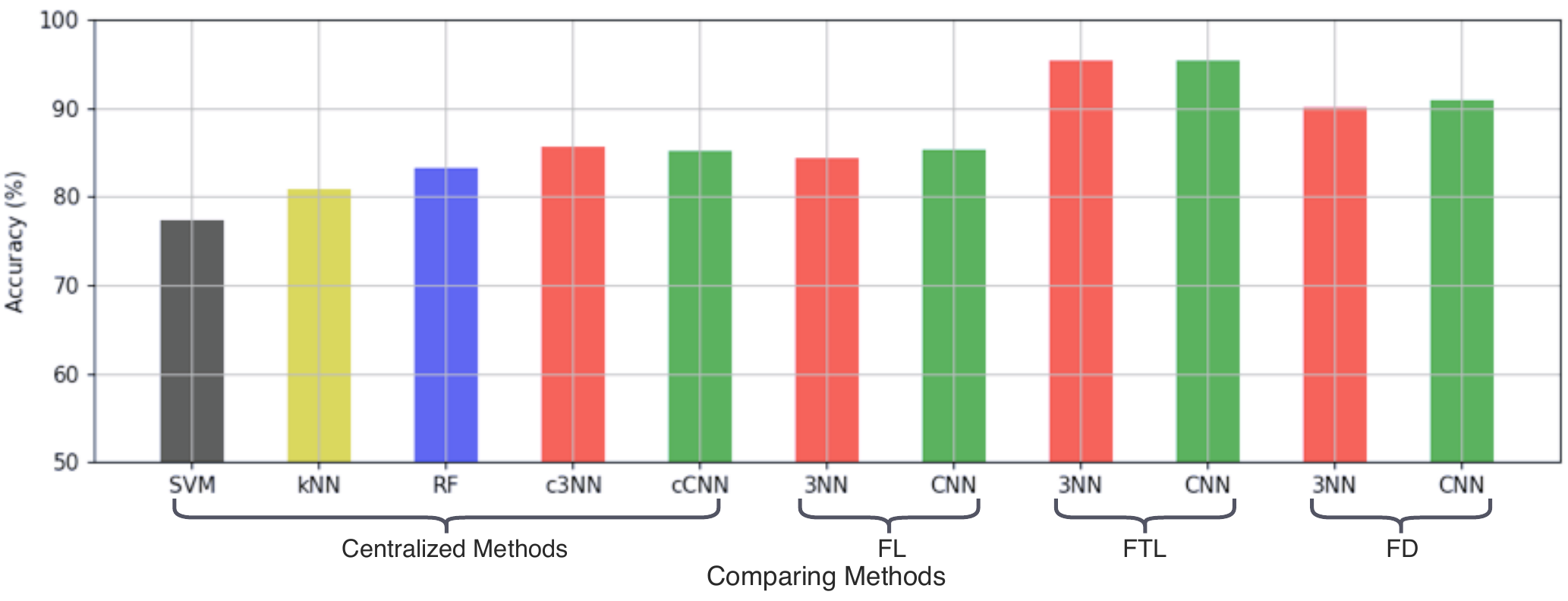}
	\caption{The accuracy of different learning methods in human activity recognition}
	\label{CM}
\end{figure*}

\begin{figure}[tbp]
	\centering
	\includegraphics[width=0.95\linewidth]{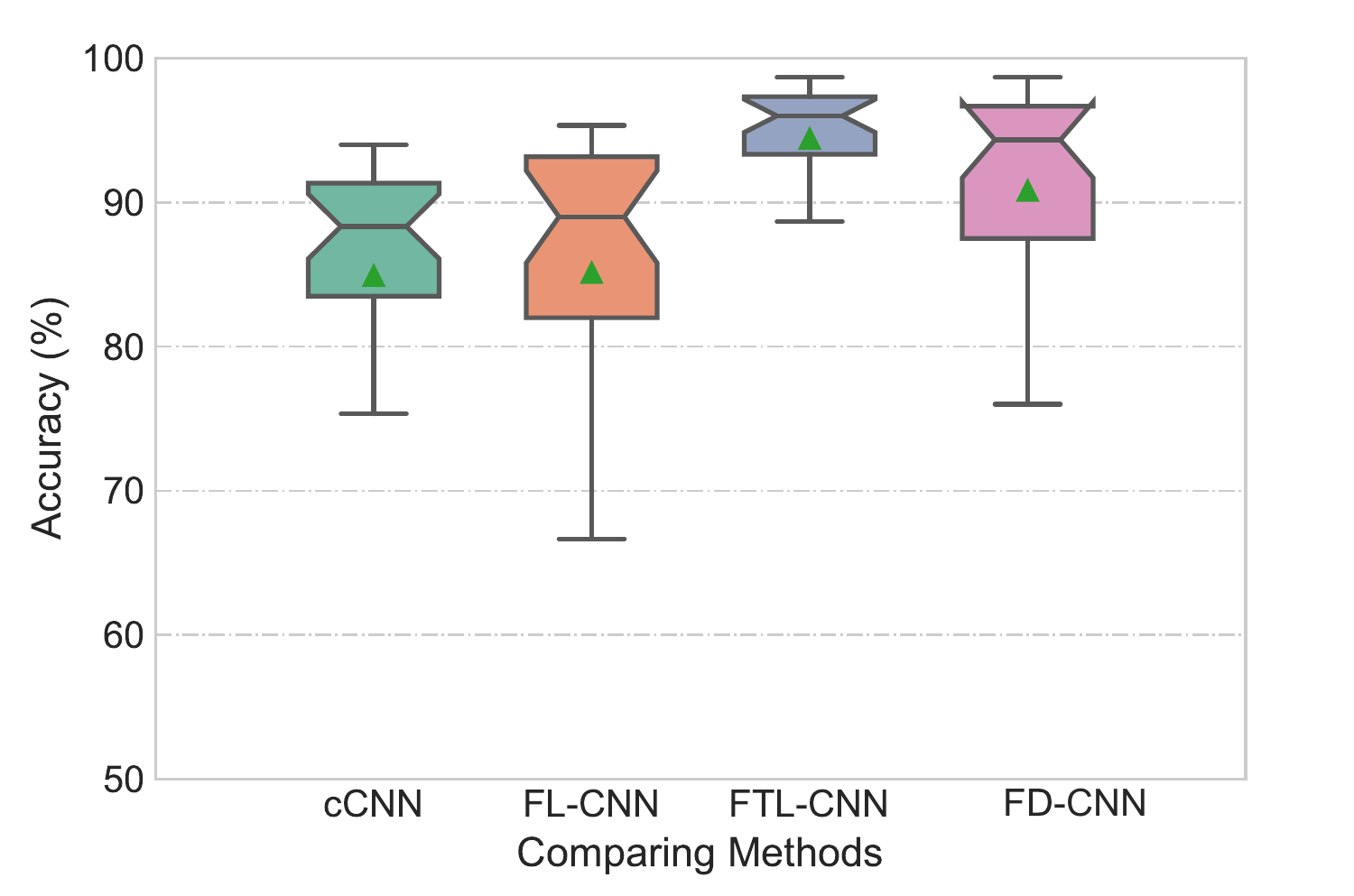}
	\caption{The accuracy distribution of different clients predicted by CNN under different learning schemes.}
	\label{fig5}
\end{figure}

Fig. \ref{CM} illustrates the test accuracy of 30 clients under different learning approaches. For centralized methods, deep learning based methods (c3NN, cCNN) can all achieve a high accuracy than traditional machine learning based methods (SVM, kNN and RF). Under the coordination of a central cloud server, the edge clients in traditional federated learning (FL-CNN) are able to collectively reap the benefits of each other's information without compromising data privacy and achieve a competitive average accuracy of $85.22\%$ similar to cCNN. The slight performance degradation in FL-3NN and FL-CNN compared with the centralized fashion results from the statistical heterogeneity inherent in federated learning settings. With personalized federated learning, both FTL and FD can capture user's fine-grained personal information and obtain a personalized model for each participant, leading to a higher test accuracy. For example, FTL-3NN can reach $95.37\%$ accuracy, which is $11.12\%$ higher than that of FL-3NN.

Furthermore, we take a more detailed observation to evaluate the performance of personalized federated learning. As shown in Fig. \ref{fig5}, we adopt boxplot to graphically depict the six-number summary of the accuracies of 30 paticipating users, which consists of the smallest observation, lower quartile, median, upper quartile, largest observation and the mean represented by green triangle. We can see that although the average performance of FL-CNN is similar with cCNN, the global model trained by FL may perform poorly on some clients. For example, the accuracy of some clients may be lower than $70\%$ while some clients can reach a high accuracy of more than $95\%$. With personalization performed by each client with its own data, the accuracies of 30 clients vary in a very small scale which indicates that personalization can significantly reduce the performance degradation caused by non-IID distribution. FD-CNN approach has an accuracy improvement of $5.69\%$ compared with FL-CNN and the performance differences between different clients have also been narrowed. This observatiion indicates that PFL can benefit most of the participating clients and thus will encourage user engagement.

\begin{figure}[tbp]
	\centering
	\includegraphics[width=0.95\linewidth]{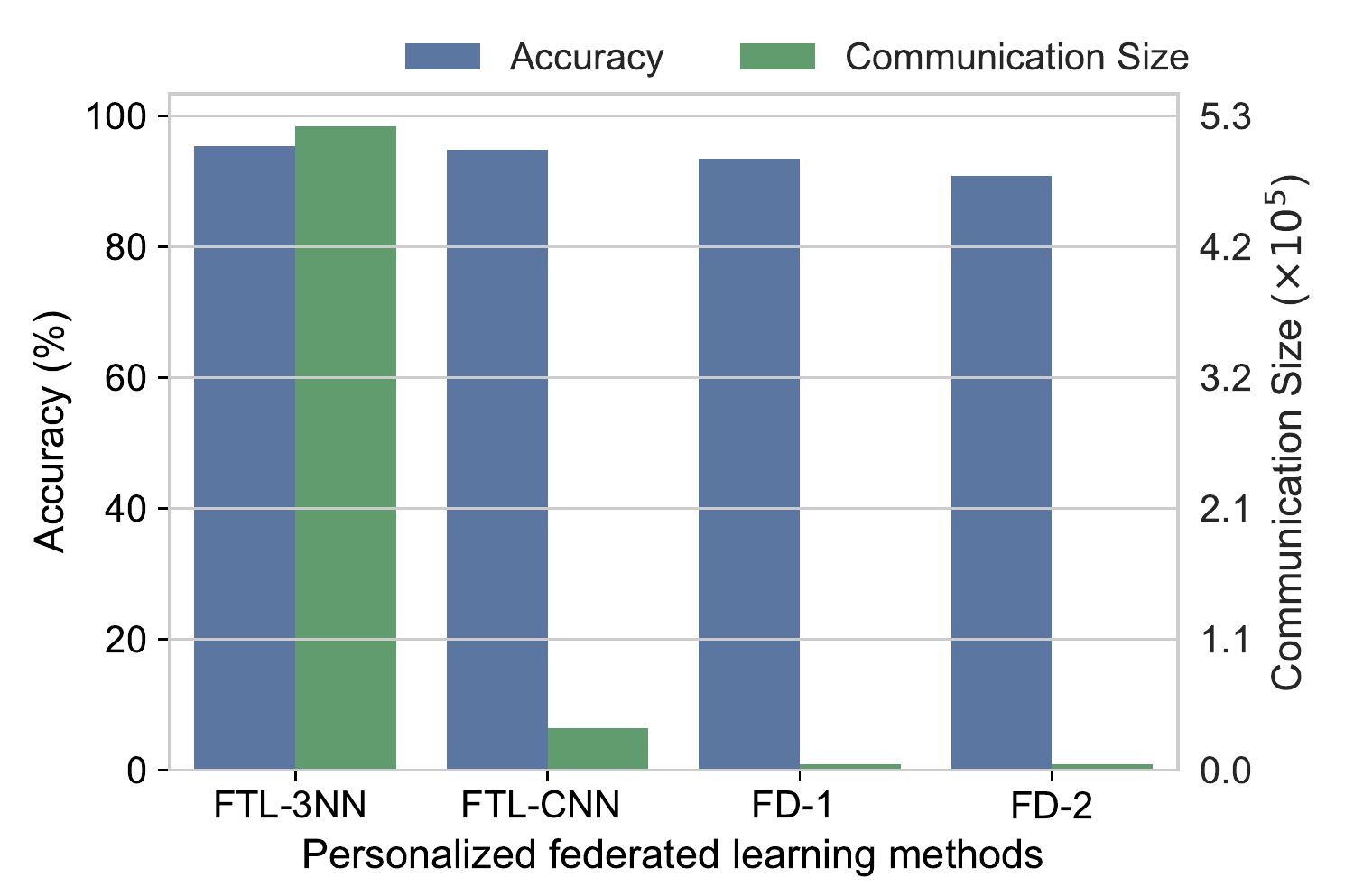}
	\caption{The accuracy and communication size of different implementations for federated transfer learning and federated distillation.}
	\label{fig6}
\end{figure}

The critical nature of communication constraints in cloud-edge scenarios also needs to be considered in federated setting because of limited bandwidth, slow and expensive connections. We compare both the accuracy and communication data size of different training models for FTL and FD. In FTL-3NN and FTL-CNN, we utilize 3NN and CNN as the model trained on both the cloud and the edge clients, respectively. For federated distillation, we consider two cases of model heterogeneity: (1) FD-1: 10 clients choose 3NN as their local models while the remaining 20 clients choose CNN; (2) FD-2: the local models of 20 clients are 3NN and the models for remaining 10 clients are CNN. As depicted in Fig. \ref{fig6}, all the four personalized federated learning methods can achieve a high accuracy of more than $90\%$. However, the communication sizes vary dramatically. As all these methods can converge within hundreds of communication rounds, we only compare the communication size in each communication round. The commnication payload size for FTL depends on the model parameter number which are 521,510 and 33,698 for FTL-3NN and FTL-CNN, respectively. While the communication size for FD is proportional to the output dimension which is 10 in our human activity recognition task. In each communication round, we randomly select 500 samples from the globally-shared data and transmit the outputted class scores predicted by each participating device to the cloud server, thus the communication size for both FD-1 and FD-2 is 5000. Fig. \ref{fig6} states that we are able to achieve superior prediction performance with lightweight models and small communication overhead, which is of great significance for supporting large-scale intelligent IoT applications.

\section{Conclusion}
In this paper, we propose PerFit, a personalized federated learning framework in a cloud-edge architecture for intelligent IoT applications with data privacy protection. PerFit enables to learn a globally-shared model by aggregating local updates from distributed IoT devices and leveraging the merits of edge computing. To tackle the device, statistical, and model heterogeneities in IoT environments, PerFit can naturally integrate a variety of personalized federated learning methods and thus achieve personalization and enhanced performance for devices in IoT applications. We demonstrate the effectiveness of PerFit through a case study of human activity recognition task, which corroborates that PerFit can be a promising approach for enabling many intelligent IoT applications.

\end{document}